\documentclass[journal=jctc,manuscript=article, layout=twocolumn]{achemso}
\usepackage[version=3]{mhchem} 
\usepackage{xcolor}
\usepackage{amssymb}
\usepackage{algorithm}
\usepackage{algpseudocode}
\usepackage{cancel}
\usepackage{url}
\usepackage{dcolumn}
\usepackage[normalem]{ulem}
\usepackage{times}
\usepackage{mathptmx}
\usepackage[pdf]{pstricks}
\usepackage{pst-all}
\usepackage{pstricks-add}

\def\CC{{C\nolinebreak[4]\hspace{-.05em}\raisebox{.4ex}{\tiny\bf ++}}}
\newcolumntype{.}{D{.}{.}{-1}}
\newcommand*{\tbc}[2]{\multicolumn{#1}{c}{#2}}

\author{Himadri Pathak}
\email{himadri.pathak@pnnl.gov}
\affiliation{Advanced Computing, Mathematics, and Data Division, Pacific Northwest National Laboratory, Richland, Washington 99354, USA}
\author{Ajay Panyala}
\affiliation{Advanced Computing, Mathematics, and Data Division, Pacific Northwest National Laboratory, Richland, Washington 99354, USA}

\author{Bo Peng}
\affiliation{Physical Sciences Division, Pacific Northwest National Laboratory, Richland, Washington 99354, United States}

\author{Nicholas P. Bauman}
\affiliation{Physical Sciences Division, Pacific Northwest National Laboratory, Richland, Washington 99354, United States}

\author{Erdal Mutlu}
\affiliation{Advanced Computing, Mathematics, and Data Division, Pacific Northwest National Laboratory, Richland, Washington 99354, USA}

\author{John J. Rehr}
\affiliation{Department of Physics, University of Washington,
Seattle, Washington 98195, United States}

\author{Fernando D. Vila}
\email{fdv@uw.edu}
\affiliation{Department of Physics, University of Washington,
Seattle, Washington 98195, United States}

\author{Karol Kowalski}
\email{karol.kowalski@pnnl.gov}
\affiliation{Physical Sciences Division, Pacific Northwest National Laboratory, Richland, Washington 99354, United States}
\title[TAMM Implementation of RT-EOM-CCSD]
{Real-time Equation-of-Motion Coupled-Cluster Cumulant Green's Function Method: Heterogeneous Parallel Implementation Based on the Tensor Algebra for Many-body Methods Infrastructure}

\abbreviations{IR,NMR,UV}
\keywords{American Chemical Society, \LaTeX}

\begin{document}

\begin{tocentry}
\includegraphics[width=3.25in]{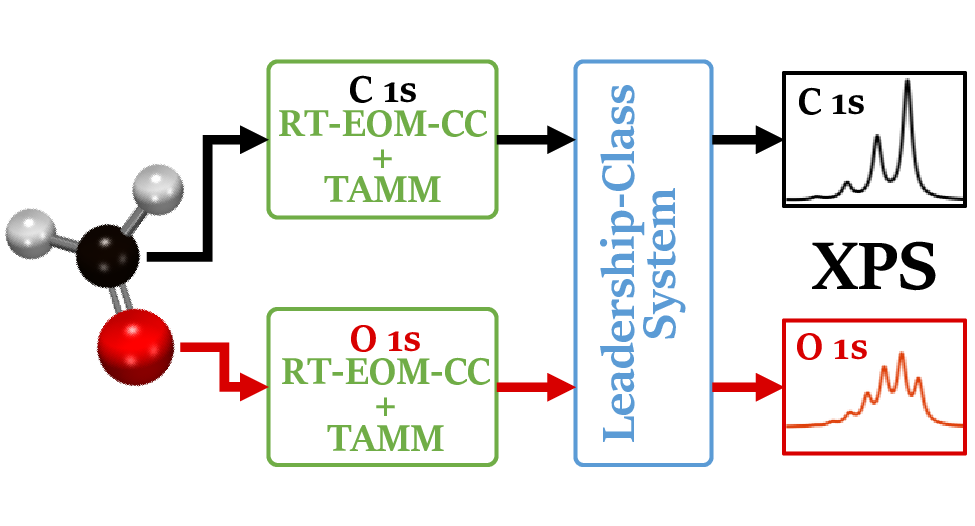}
For Table of Contents Only
\end{tocentry}    

\begin{abstract}
We report the implementation of the real-time equation-of-motion coupled-cluster (RT-EOM-CC) cumulant Green's function method [J. Chem. Phys. {\bf 152}, 174113 (2020)]
within the Tensor Algebra for Many-body Methods (TAMM) infrastructure. TAMM is a massively parallel heterogeneous tensor library designed for utilizing forthcoming exascale computing resources. The two-body electron repulsion matrix elements are Cholesky-decomposed, and we imposed spin-explicit forms 
of the various operators when evaluating the tensor contractions. 
Unlike our previous real algebra Tensor Contraction Engine (TCE) implementation, the TAMM implementation supports fully complex algebra.
The RT-EOM-CC singles (S) and doubles (D) time-dependent amplitudes are propagated using a first-order Adams--Moulton method. This new implementation shows excellent scalability tested up to 500 GPUs using the Zn-porphyrin molecule with 655 basis functions, with parallel efficiencies above 90\% up to 400 GPUs. The TAMM RT-EOM-CCSD was used to study core photo-emission spectra in the formaldehyde and ethyl trifluoroacetate (ESCA) molecules. Simulations of the latter involve as many as 71 occupied and 649 virtual orbitals. The relative quasiparticle ionization energies and overall spectral functions agree well with available experimental results.
\end{abstract}

\section{Introduction\label{sec:sec1}}
Photoemission spectroscopy (PES) is a widely used spectroscopic probe, covering a broad energy range from a few to several thousands of electronvolts (eV).
The ubiquity of this technique is due, in part, to the variety of instruments available, ranging from small laboratory-based ones to synchrotron facilities.\cite{meirer2018spatial,D1TA08254J}
In the UV energy regime, PES provides access to the valence electronic structure, where low-energy electrons are the driving force of many chemical and biological processes.\cite{C8CP07831A} In the X-ray regime, core or X-ray photoelectron spectroscopy (XPS) is one of the most commonly used fingerprinting method in materials science, catalysis, and chemical engineering, where it is used to investigate the composition and chemistry of materials at the atomic level.\cite{seah1980quantitative,bagus2013interpretation,shchukarev2006xps} 
Given the importance of this experimental technique,   complementary developments of accurate theoretical methods and associated software for the calculation and interpretation of XPS are crucial. Thus, a variety of methods have been developed in both the frequency and time domains.  They are often used to predict the position of the main transition or quasiparticle (QP) peak. However, only advanced electron correlation methods can accurately simulate details of core-level photoemission spectra, particularly for the shake-up peaks, since these many-body phenomena reflect a complex interplay between electron correlation and orbital-reorganization effects.
The time-independent coupled-cluster (CC) method and its many extensions \cite{coester58_421, coester60_477, cizek66_4256, paldus72_50, kummel2003biography, RevModPhys.79.291, mukherjee1989use, krylov2008equation, piecuch2002recent, crawford2000introduction} have proven their utility in recovering electron correlation in a variety of problems for both ground and excited states. The ground-state CC 
method is size-extensive 
at any level of truncation of the excitation operators, and scales polynomially with the number of active orbitals.
This makes these methods an attractive choice over other electron correlation methods, as they provide a balanced trade-off between the computation cost and desired accuracy. Furthermore, it is possible to  improve the results systematically by incorporating more correlated determinantal spaces. The key feature of CC methods is the use of an exponential parameterization of the correlated ground state wavefunction $\left|\Psi\right>$: $\left|\Psi\right>=e^{T}\left|\Phi\right>$, where $T$ is the cluster operator, and $|\Phi\rangle$ is a reference wavefunction, which is usually  but not necessarily a Hartree--Fock wavefunction. The cluster operator is defined by order of excitation, i.e. $T = T_1+T_2+\dots+T_n$ corresponding to singles, doubles, triples, ..., $n$-tuples, 
where
$T_n=(\frac{1}{n!})^2 \sum_{\substack{i_1,\dots,i_n\\a_1,\dots,a_n}} t_{i_1\dots i_n}^{a_1\dots a_n}\,\,a^\dag_{a_1}\dots a^\dag_{a_n}a_{i_n}\dots a_{i_i}$.
Here $a_p^\dag$ and $a_p$ are creation and annihilation operators, respectively, associated with a set of $N_{so} = 2N_{bas}$ orthonormal spin-orbitals
$\left\{\phi_p \right\}$, with $N_{bas}$ being the number of basis functions. The indices $i_n$ ($a_n$) correspond to orbitals that are occupied (unoccupied) with respect to the reference determinant.

When combined with the Green's function (GF) formalism in the frequency domain, the CC method   provides an avenue to treat excited-state correlation effects that play a crucial role in accurately simulating quasiparticles and satellite peaks in XPS.\cite{nooijen1992coupled, nooijen1993coupled, nooijen1995second, meissner1993electron, bhaskaran2016coupled, peng2016coupled, peng2018properties, peng2021coupled, shee2019coupled, lange2018relation} 
However, for large systems, time-domain methods offer  advantages over their frequency-domain counterparts by trading off memory resources for the serialization of the calculation. Therefore, there has been a lot of effort in developing efficient time-dependent approaches.
Hoodbhoy and Negele\cite{hoodbhoy1978time, hoodbhoy1979time}, and Sch{\"o}nhammer and Gunnarsson\cite{schonhammer1978time} reported  formulations of a time-dependent CC theory at about the same time. More recently, Kvaal proposed an orbital-adaptive time-dependent coupled-cluster method \cite{kvaal2012ab} relying on Arponen's bi-orthogonal formulation of CC theory,\cite{arponen1983variational} considering the complex analytic action formulation of the time-dependent variational principle (TDVP). Sato \textit{et al.} have also developed a time-dependent optimized coupled-cluster method considering the real-action formulation of the TDVP,\cite{sato2018communication, pathak2020time} to approximately solve the time-dependent Schr{\"o}dinger equation (TDSE) as a polynomial cost-scaling alternative to 
multiconfiguration time-dependent Hartree--Fock
methods.\cite{caillat2005correlated, kato2004time, sato2013time} Both approaches\cite{kvaal2012ab, sato2018communication, pathak2020time} choose to optimize the orbitals and ignore the one-body excitation ($T_1$) and de-excitation ($\Lambda_1$) operators. 
This approximation is well suited for strong-field physics, where consideration of optimal orbitals is  crucial to obtain meaningful results. Such approaches provide a gauge-invariant description of the time-dependent properties of interest and satisfy the Ehrenfest theorem due to the use of variationally optimized orbitals.
\cite{ishikawa2015review, sato2022time, pathak2020time2, pathak2020study}.   
Despite this advantage, such methods\cite{kvaal2012ab, sato2018communication, pathak2020time} are ill-suited for large-scale applications, especially when simulations involve core-hole states of chemical systems containing many active occupied electrons. Other developments in time-dependent electronic structure theory\cite{li2020real} and time-dependent coupled-cluster methods\cite{monkhorst77_421, mukherjee1979response, guha1991multireference, dalgaard1983some, koch1990coupled, takahashi1986time, prasad1988time, sebastian1985correlation, pigg2012time, nascimento2016linear, nascimento2017simulation, koulias2019relativistic, nascimento2019general, pathak2020time2, pathak2021time,  park2019equation, huber2011explicitly, shushkov2019real, ChanWhite, pedersen2019symplectic} are reviewed elsewhere.

Recently, we have developed a real-time equation-of-motion coupled-cluster (RT-EOM-CC) cumulant Green’s function method\cite{rehr2020equation,vila2020real, vila2021equation, vila2022real, vila2022_water} building on the Sch{\"o}nhammer and Gunnarsson formulation of the TDCC\cite{schonhammer1978time}. Subsequently several applications to the XPS of small molecules containing a few electrons in moderate size basis have been reported.\cite{vila2020real, vila2021equation, vila2022real, vila2022_water}
In this methodology, as described in more detail below, the Green's function has a natural exponential cumulant form, which is given by solutions to a set of coupled, first-order, nonlinear   differential equations for the time-dependent CC amplitudes.   While the traditional cumulant approximation is linear in the one-particle self-energy, the RT-EOM-CC approach builds in  high-order nonlinear, nonperturbative contributions.

Even with their inherent memory usage advantage, large-scale time-dependent simulations are computationally challenging   for chemically relevant systems that go beyond a handful of correlated electrons. However, thanks to recent advances in high-performance computing techniques that can take advantage of peta- and eventually exascale computational resources, such calculations are no longer insurmountable. In this article, we report the implementation of the RT-EOM-CC method with single and double excitations (RT-EOM-CCSD) within the Tensor Algebra for Many-body Methods (TAMM) infrastructure\cite{mutlu2022tamm, mutlu2019toward}. TAMM is a massively parallel heterogeneous tensor library designed for developing quantum chemistry applications for forthcoming exascale supercomputers. Our RT-EOM-CC code uses Cholesky-decomposed two-electron repulsion matrix elements\cite{bo1, bo2} that aid in reducing the memory requirements and inter-node communication. Other speed-ups arise from spin-explicit evaluation of the coupled-cluster amplitudes. 
As discussed in the next section, the CC amplitudes in RT-EOM-CC are naturally complex valued. In contrast, our original implementation based on the real-valued Tensor Contraction Engine (TCE)\cite{vila2020real, vila2022real} used separate real-valued data structures to represent the real and imaginary parts and required two distinct subroutines to handle the propagation. Since the new TAMM implementation uses explicit complex algebra,   only   a single subroutine is required to handle the complex data, thus reducing the coding and data intricacy, even though complex algebra is more floating point operation intensive.

To demonstrate the capabilities of this new implementation, we study the core spectral functions of the formaldehyde and ethyl trifluoroacetate (ESCA) molecules and compare them with experimental spectra. We observe satisfying agreement between the computed and available experimental spectra.   
In addition to these systems, we present parallel and storage performance for a few nominally ``large'' systems such as Zn-porphyrins, uracil, and the benzene-ammonia dimer.

\section{Methods}
\subsection{Real-time Equation-of-Motion\\ Coupled-Cluster Cumulant Green's Function Method}

The many-body Green's function approach has proved very useful for the calculation of spectral functions of extended systems.\cite{PhysRevB.90.085112,PhysRevB.91.121112,PhysRevB.94.035156,doi:10.1116/6.0001173} By combining this approach with the CC method, we have developed  a time-dependent CC cumulant Green's function method\cite{rehr2020equation} that integrates the advantages of both.
A detailed derivation of the complete method can be found elsewhere.\cite{vila2020real, vila2021equation, vila2022real, vila2022_water}. In this section, we give a brief introduction to the RT-EOM-CC formulation.
The goal is to construct the retarded core-hole Green's function $G_c^R(\omega)$
and associated core spectral function $A_c(\omega)=(-1/\pi)\,\mathrm{Im}\, G_c^R(\omega)$,
by introducing a time-dependent coupled-cluster ansatz, $e^{iH\tau}\left|\Psi\right>=\left|\Psi(\tau)\right> = N_c(\tau)e^{T(\tau)}\left|\Phi\right>$, which is a formal solution to the time-dependent Schr{\"o}dinger equation 
$-i \frac{\partial}{\partial \tau}\left|\Psi(\tau)\right>= H
\left|\Psi(\tau)\right>$.
Here $\tau$ is time, and $\left|\Psi(\tau)\right>$ is the fully correlated wavefunction for the $(N-1)$-electron state,
$H = \sum_{pq} h_{pq} a^\dag_p a_q + \frac{1}{4} \sum_{pqrs}v_{pq}^{rs}a^\dag_p a^\dag_q a_s a_r$ is the nonrelativistic electronic Hamiltonian in second-quantization form, $h_{pq}$ are the single-particle kinetic and electron-nuclei spinorbital integrals, and $v_{pq}^{rs} = \left<pq\left|\right|rs\right>$ are the usual antisymmetrized two-particle Coulomb integrals.
$N_c(\tau)$ and $T(\tau)$ are, respectively, the time-dependent normalization constant and the time-dependent coupled-cluster operator. In principle, $c$ can be any occupied orbital, but here we focus on deep core excitations. As usual, the
$p,q,r,s$ indices indicate generic spin-orbital states. In the current formulation, the time-independent reference determinant $\left|\Phi\right>$ is a single $(N-1)$-electron determinant formed from the $N$-electron Hartree--Fock states where the state $c$ has been annihilated. Thus, it is important to note that $T$ acts in the $(N-1)$-electron space where $c$ is now included in the set of unoccupied single-particle states. Following Ref.~\citenum{rehr2020equation},  the final form of the equations of motion (EOMs) for $N_c(\tau)$, and $t_{ij\dots}^{ab\dots}(\tau)$ are:
\begin{eqnarray}
&&-i\,\partial_\tau\,\mathrm{ln} N_c(\tau)= 
\langle \Phi|(H_Ne^{T(\tau)})_C|\Phi\rangle+E_{N-1}^{\mathrm{HF}} \label{eqn:N_c}\,\,\,\,\,\,\,\,\,\\
&&-i\,\partial_\tau\,t_{ij\dots}^{ab\dots}(\tau)\,=\langle \Phi_{ij\dots}^{ab\dots}|(H_Ne^{T(\tau)})_C|\Phi\rangle \label{eqn:amp},
\end{eqnarray}
Here 
$H_N = \sum_{pq} f_{pq} \{a^\dag_p a_q\}' + \frac{1}{4} \sum_{pqrs} v_{pq}^{rs} \{a^\dag_p a^\dag_q a_s a_r\}'$, and $\{\}'$ indicates that the normal ordering is done with respect to $\left|\Phi\right>$ instead of the usual $N$-electron Hartree--Fock determinant. $E_{N-1}^{\mathrm{HF}}=\left<\Phi\left|H\right|\Phi\right>$ is the Hartree--Fock energy of the $(N-1)$-electron systems. The Fock operator matrix elements are $f_{pq}= \epsilon_p \delta_{pq}-v_{pc}^{qc}$, where $c$ is the core-hole index as used above.
The subscript "C" designates a connected part of a given operator expression.



Within this CC approximation, the retarded core-hole GF $G_c^R(\tau)$ is simply proportional to the normalization factor $N_c(\tau)$, since Eq.~\ref{eqn:N_c} has an exponential solution, and therefore $C_c^R(\tau)$, the retarded cumulant associated with $c$, is proportional to $\mathrm{ln}N_c(\tau)$:
\begin{eqnarray}
\label{eq:cum_gf}
G_c^R(\tau)
&=&-i\Theta(\tau) e^{-i (\epsilon_c + E_N^{corr}) \tau} N_c(\tau)\\
\label{eq:cum_gf2}
&=& -i  \Theta(\tau) e^{-i (\epsilon_c + E_N^{corr}) \tau} e^{C_c^{R}(\tau)}\label{eq:cum_gf}\\
C_c^R(\tau)&=&i\int_0^\tau\left<\Phi\left|(H_Ne^{T(\tau')})_C\right|\Phi\right> d\tau'.
\label{eq:cum_gf3}
\end{eqnarray}
Here $\epsilon_c$ is the single-particle Hartree--Fock energy of the core orbital $c$, and E$_N^{corr}$ corresponds to the correlation energy of the $N$-electron closed-shell ground state. As a consequence, $G_c^R(\tau)$ can be expressed as the product of the free particle GF and the exponential of a cumulant, as expected within the cumulant approximation.    
We can now write down from Eq.~\ref{eqn:N_c} a complete expression for the time derivative of $C_c^R(\tau)$ within the single and double excitations approximation (CCSD):
\begin{equation}
\label{eq:dcdt}
\begin{split}
-i{\partial_\tau C_c^R}&=\left< \Phi\left|\left[{H}_N, T_2(\tau)\right]\right|\Phi\right>\\
&+\frac{1}{2}\left< \Phi\left|\left[\left[{H}_N, T_1(\tau)\right], T_1(\tau)\right]\right|\Phi\right>
\end{split}
\end{equation}

Unlike conventional linear self-energy formulations, it is evident from Eq. \ref{eq:dcdt} that the CCSD cumulant includes nonlinear, nonperturbative contributions.

The EOMs for the amplitudes within the CCSD approximation are obtained from Eq. \ref{eqn:amp} as
\noindent
\begin{align}
-i\partial_\tau t_i^a(\tau)
    &=\left < \Phi_i^a|(H_Ne^{T_1(\tau)+T_2(\tau)})_C|\Phi\right > \label{eq:dt1eom}\\
-i\partial_\tau t_{ij}^{ab} (\tau))
    &=\left < \Phi_{ij}^{ab}|(H_Ne^{T_1(\tau)+T_2(\tau)})_C|\Phi\right>\;,\label{eq:dt2eom}
\end{align}
where $t_i^a(\tau)$ and $t_{ij}^{ab} (\tau))$ are time-dependent singly and doubly excited cluster amplitudes.



\begin{figure}[t]
\includegraphics[width=3.33in]{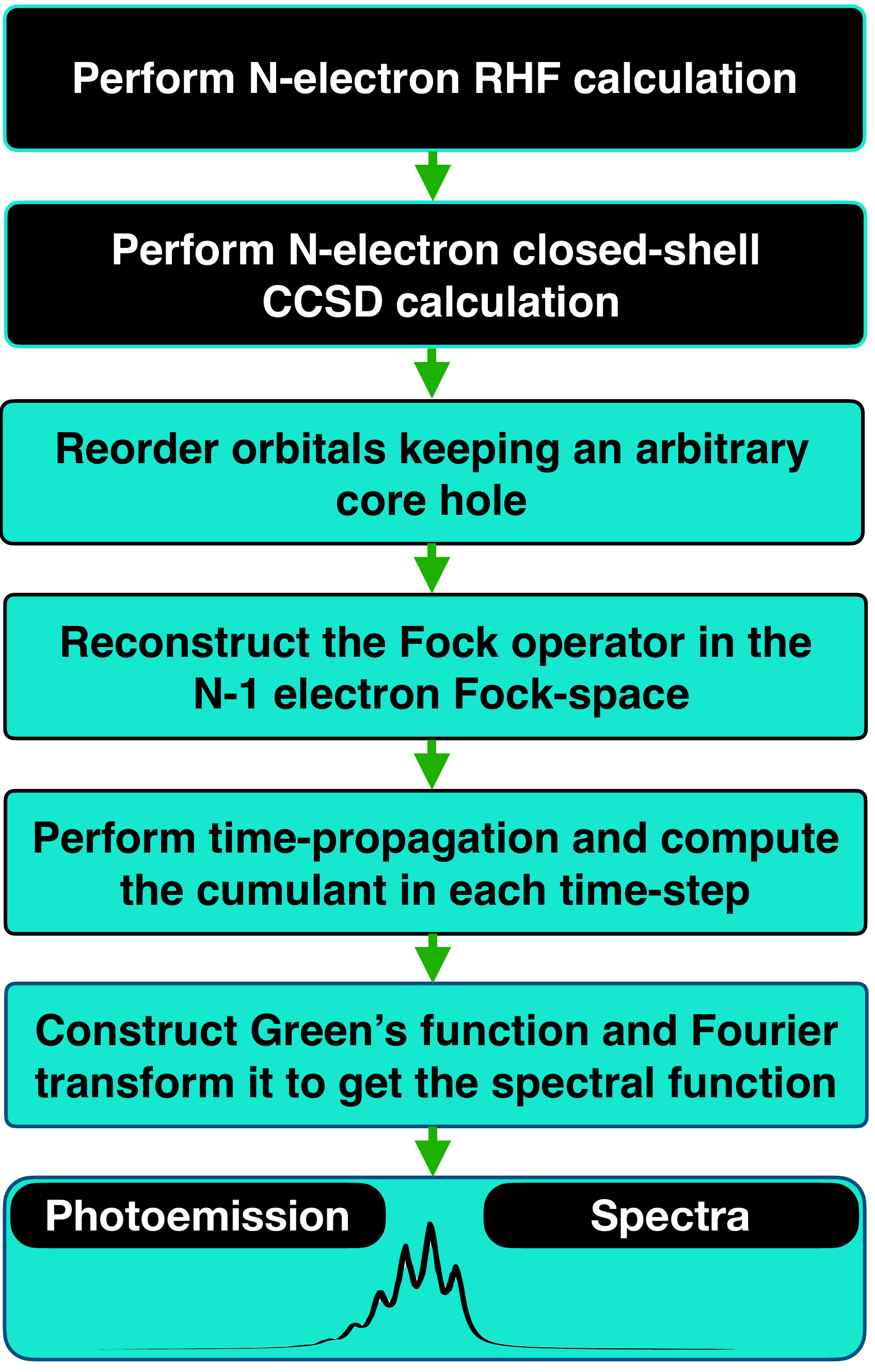}
\caption{Real-time equation-of-motion coupled-cluster workflow.}
\label{fig:rteomcc_workflow}
\end{figure}

\section{Numerical Solution of the EOMs}
To construct the core-hole spectral function $A_c(\omega)$ we need to compute the Green's function $G_c^R(\tau)$, which in turn depends on the cumulant $C_c^R(\tau)$ over the whole simulation range (Eqs.~\ref{eq:cum_gf3} and \ref{eq:dcdt}). Thus, the main task is to propagate Eqs.~\ref{eq:dt1eom} and \ref{eq:dt2eom} in time, which provide solutions for both the $T_1$ and $T_2$ amplitudes needed to compute the cumulant $C_c^R(\tau)$.
Fig.~\ref{fig:rteomcc_workflow} demonstrates a typical RT-EOM-CC workflow.
Given that the  RT-EOM-CC uses ground-state orbitals and the closed-shell CC ground-state energy, stationary restricted Hartree--Fock (RHF) and closed-shell coupled-cluster calculations are prerequisites to the time-dependent simulation.
The computational cost of the RHF calculations is $\mathcal{O}(N_{bas}^3)$ while that for the   CC part varies, depending on the imposed truncation in the excitation operator. The singles and doubles approximation used in this work scales as $\mathcal{O}(O^2V^4)$, where $O$ and $V$ denote the total number of occupied and virtual orbitals, respectively, and $O+V=N_{bas}$.

The first two ground-state calculation steps are common to most simulations involving   core-hole excitations, but these  are not the main bottlenecks since they need to be performed only once.  After these, all calculations involving different core-hole states are unique and can run simultaneously.
These calculations start by reading the $N$-electron closed-shell orbitals into the $(N-1)$-electron orbital space
and then constructing the $(N-1)$-electron Fock matrix.
At this point, the time-dependent simulation can start, where we propagate the CCSD amplitudes forward for the desired time, computing the time-derivative of the cumulant at each time step. It is worth pointing out that given that the time-dependent simulations involve an initial core hole, this is an open-shell problem and is roughly three times more expensive than its closed-shell counterpart in spin-explicit form. 

As in our previous implementations, the first-order coupled nonlinear simultaneous differential equations for the amplitudes are integrated using the first-order Adams--Moulton method,\cite{quarteroni2010numerical} also known as the implicit trapezoidal rule. 
In this approximation the $t_{ij\dots}^{ab\dots}(\tau)$ are propagated with:
\begin{equation}
\label{eq:trapzrule}
\begin{split}
t_{ij\dots}^{ab\dots}(\tau+\Delta \tau)&=~t_{ij\dots}^{ab\dots}(\tau)\\
&+\frac{i}{2}\Delta \tau\Big(\left<\Phi_{ij\dots}^{ab\dots}\left| (H_N e^{T(\tau)})_C\right|\Phi\right>\\
&+\left< \Phi_{ij\dots}^{ab\dots}\left|
(H_N e^{T(\tau+\Delta\tau)})_C
\right|\Phi\right> \Big),
\end{split}
\end{equation}
where $\Delta \tau$ is the simulation time-step. The propagation starts with the initial conditions $T_1=0$ and $T_2=0$. Since the Adams--Moulton method is implicit, the values of the amplitudes at $(\tau+\Delta \tau)$ depend on themselves. Thus, to solve Eq. \ref{eq:trapzrule} we use a fixed-point iteration scheme at each time step. Other methods of solution that use a variable time step and are more stable than fixed-point iteration are currently under development.
After completion of the time propagation, the remainder of the workflow is not compute-intensive, since it only involves forming the time-dependent Green's function from the cumulant, and then Fourier-transforming  to obtain the spectral function in the frequency domain.

\subsection{RT-EOM-CC Implementation in Tensor Algebra for Many-body Methods Infrastructure}
\begin{figure}[t]
\includegraphics[width=3.33in]{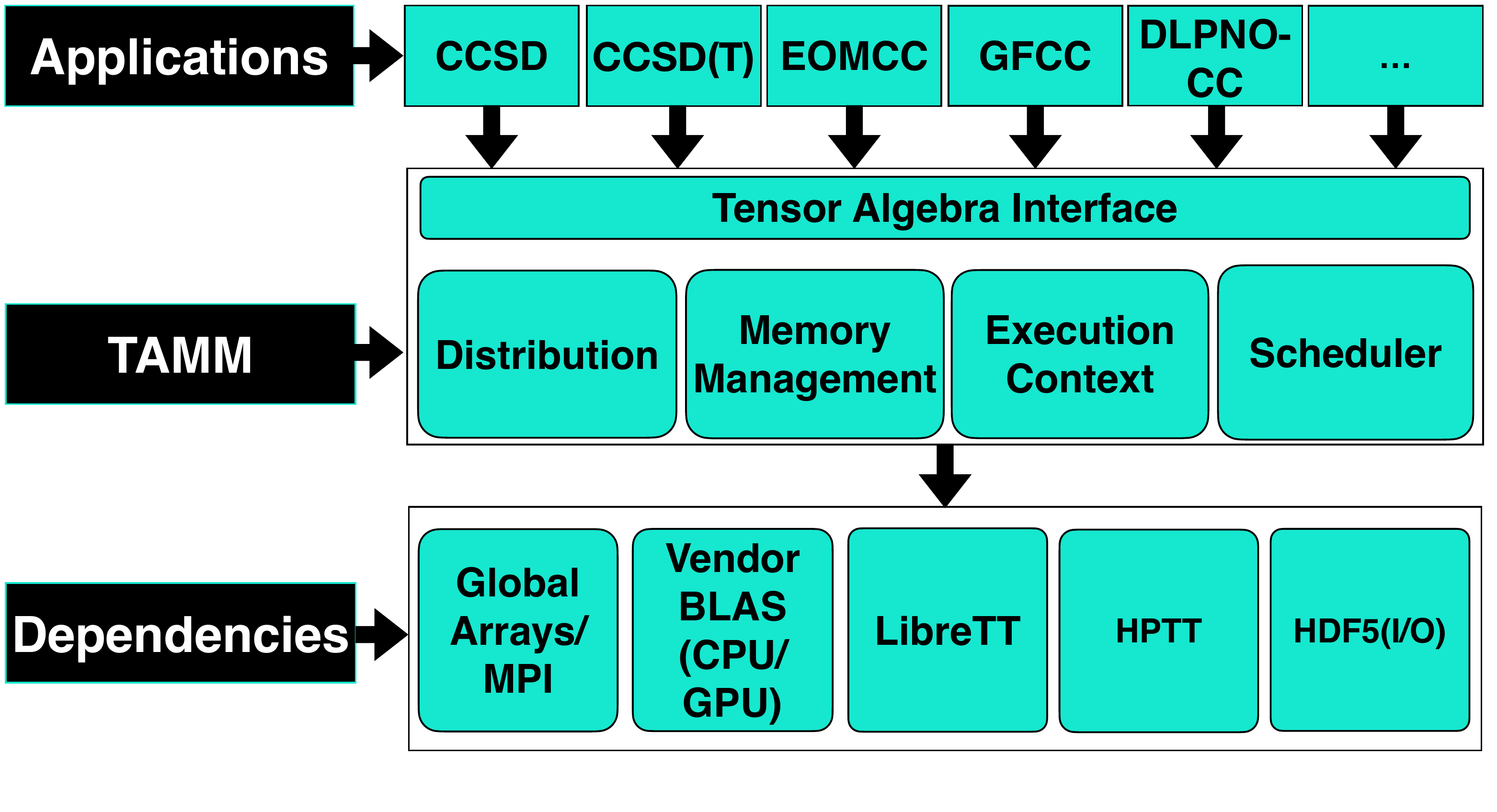}
\caption{Overview of the Tensor Algebra for Many-body Methods (TAMM) framework.}
\label{fig:Tamm_fig}
\end{figure}

\begin{table*}[t]
\centering
\caption{Structural features of TAMM, including various used third-party dependency libraries.}
\label{tab:software_lang}
\begin{tabular}{ll}
\hline
Third-party dependencies & Global Arrays~\cite{ga_paper}, BLIS~\cite{BLIS1}, vendor BLAS/LAPACK, \\
& cuBLAS/rocBLAS/oneMKL \\
                          & HPTT~\cite{hptt_17}, TALSH~\cite{talsh}, LibInt2~\cite{libint} \\
\hline
 Programming Languages & \CC17, CUDA, HIP, SYCL, MPI, OpenMP \\
\hline
 Precision & Double \\
 \hline
 Data Types & Real, Complex \\
 \hline
 Supports Restart Capabilities? & Yes \\
 \hline
 I/O requirements & Minor \\
 \hline
\end{tabular}
\end{table*}

The Tensor Algebra for Many-body Methods (TAMM) library \cite{mutlu2022tamm,kowalski2021nwchem} is a massively parallel, heterogeneous tensor algebra library. It provides a computational infrastructure (see Fig.~\ref{fig:Tamm_fig}) that can achieve scalable performance. Furthermore, it allows portable implementations of many-body methods both on existing and forthcoming exascale super-computing platforms.

High-dimensional tensor contractions are the most compute-intensive components of the RT-EOM-CC method. Within the single and double excitations approximation (RT-EOM-CCSD), the most expensive tensor contraction is of the form
$R(V, V, O, O)  = \alpha\times v(V, V, V, V)\times t_2(V, V, O, O)$,
involving four-dimensional tensors, the antisymmetrized two-body matrix elements $v$, and the two-body CC excitation operator $t_2$, 
$\alpha$ is a scalar pre-factor.
These multi-dimensional tensor contractions are not only computational- but also communication-intensive.
There have been many efforts to develop specialized parallel tensor algebra libraries, including automated code generators and better memory management to facilitate these demanding tensor contractions.\cite{flame, solomonik2014massively, epifanovsky2013new, ibrahim2014analysis} 
The TAMM library is one such effort aiming to achieve scalable performance on several heterogeneous architectures,  by delivering a common platform for the portable implementation of numerous many-body methods. TAMM provides a variety of features to   users including the ability to specify and manipulate tensor distribution, memory management, and scheduling of tensor operations. In addition, it supports both complex and mixed real-complex algebra for mathematical operations.

A summary of structural features and third-party libraries for the TAMM library is shown in Table~\ref{tab:software_lang}.
TAMM uses Global Arrays~\cite{ga_paper} and Message Passing Interface (MPI) to achieve scalable parallelization on distributed memory platforms, while using optimized libraries that help efficient intra-node execution of the tensor operation, both in CPU kernels and accelerators. 
 TAMM also  uses multi-granular dependence analysis and task-based execution to execute operations. First, it constructs  
a macro operation graph by analyzing the dependencies between various operations. When two operations share the same data structure, with one of them writing to the other, they are in conflict and impossible to execute in parallel. The operation graph is analyzed to identify and order the non-parallel operations to minimize the required numbers of synchronizations. The possible scheduled operations are executed in a single program multiple data fashion. Such executions are compatible with MPI, and their collective executions are  performed on a given MPI communicator.
Various tasks constitute an operation, which  is produced using task iterators. Each task performs part of the computation, usually adding a block of data to the output tensor. A given task is migratable and can be scheduled for execution on any compute node or processor core until its execution begins. The data needed for a job are transported to its location once the execution of the process has started. At this stage, migration of tasks is no longer possible and they are bound to process. 
TAMM's GPU execution scheme uses localized summation loops to minimize the transfer of output blocks from GPUs to CPUs. This helps to reduce the data transmission between CPUs and GPUs by keeping the output block that is being updated by several input tensor blocks on the GPU until all updates are complete. 

 \subsubsection{Programming Models, Software Dependencies, I/O, Restart/Checkpoint Capabilities} \label{sec:software_lang}
Memory demands, operation count, and time-to-solution are three main concerns that limit large-scale CC calculations. This is further complicated when extending the CC formalism to the time domain. To combat these hindrances, various techniques and features were incorporated during the implementation of the RT-EOM-CCSD method.

In canonical spin-orbital CC calculations, the four-dimensional electron repulsion integral (ERI) tensors are easily the largest memory-demanding objects. The storage requirement for the ERI tensor in its full spin-orbital form is of order $N_{so}^4$, and they must undergo a tensor transformation from the atomic-orbital to molecular-orbital basis which scales as $\mathcal{O}(N_{bas}^5$). In our RT-EOM-CC implementation, we employed Cholesky decomposition\cite{cd1, cd2, cd3, cd4, cd5} of the ERI tensors to ease the memory/storage demands and increase the data locality resulting in reduced communication. We leveraged the same Cholesky decomposition previously implemented using the TAMM library\cite{bo1, bo2}, which is an on-the-fly pivoting decomposition of the two-body ERIs.  The resulting Cholesky bases are three-index quantities that reduce the storage requirements from order $N_{so}^4$ to $KN_{so}^2$, where $K\sim \mathcal{O}(N_{bas})$ is the number of Cholesky bases. In addition, the atomic-orbital to molecular-orbital transformation is conducted on the Cholesky bases rather than directly on the ERI, thus reducing the scaling of the transformation from $\mathcal{O}(N_{bas}^5)$ to $\mathcal{O}(N_{bas}^4)$. The accuracy of the correlation calculations employing the Cholesky bases in comparison with the canonical results is well-controlled through adjusting the diagonal cutoff in the Cholesky decomposition. 

The next largest memory-demanding objects, after the ERIs, are the $T_2$ amplitudes. In a naive spin-orbital implementation, the memory requirement for all elements of the $T_2$ operator is $16\,O^2V^2$. The declaration includes 16 possible combinations of spins for the four indices, most of which do not contribute or are over-specified as they are equivalent through permutational symmetry. In our RT-EOMCC methodology, we implemented the spin-integrated form of equations. Only unique spin combinations of tensors with a non-zero contribution, including intermediates, are programmed. For the $T_2$ amplitudes, this means only three spin cases are necessary ($t_{\alpha\alpha}^{\alpha\alpha}$, $t_{\beta\beta}^{\beta\beta}$, and $t_{\alpha\beta}^{\alpha\beta}$), reducing the memory requirements to $\sim\frac{3}{16}$ that of the full $T_2$ operator. Memory requirements are reduced for other operators as well. Since non-contributing spin combinations are removed, and only unique spin cases are specified, the overall operation count is significantly reduced. The memory requirements of various systems with various large systems with 760 to 2450 spin-orbital functions can be seen in Table~\ref{tab:io_req}. In this Table, the memory requirement for the cluster amplitudes reflects the sum of the three timeline tensors needs for the iterative update given by Eq.~\ref{eq:trapzrule}. The combination of Cholesky-decomposed ERIs and spin-integrated equations allows for simulations with hundreds of orbitals on moderately sized computer clusters. 

Another challenging aspect of RT-EOM-CC calculations is that it is necessary to propagate for a sufficiently long time to have well-resolved spectra. In practice, shared computing resource usually do not allow simulations to propagate for enough time in a single run to achieve sufficient resolution. Furthermore, it is important to be able to track simulations and adjust/optimize the time propagation parameters promptly, before a long time has elapsed. For these reasons, checkpointing and restart capabilities were imperative in our RT-EOM-CCSD implementation. Restarting the time-propagation algorithm at any $i^{\mathrm{th}}$ step requires the $T$ amplitudes of the $(i-1)^{\mathrm{th}}$ and $i^{\mathrm{th}}$ time steps, in addition to the Fock and Cholesky-decomposed ERIs. Parallel read and write capabilities in the TAMM library allow for periodic checkpointing while minimizing their impact on the overall computational time.

\begin{table*}[t]
\caption{Workflow memory requirements for the TAMM implementation of RT-EOM-CCSD for systems with Sapporo-TZP basis set for all atoms except H, for which we use aug-cc-pVTZ, using a Cholesky vectors diagonal cutoff of $10^{-6}$, and linear dependence threshold of $10^{-6}$.}
\label{tab:io_req}
\begin{tabular}{lrrrrrrrc}
\hline
System & \tbc{5}{Configuration Space} & \tbc{1}{$T_2$} & \tbc{1}{Cholesky} & \tbc{1}{\# of Cholesky}  \\
    &     & & & &                              &  \tbc{1}{amplitude}   &  \tbc{1}{vectors}  & \tbc{1}{vectors} \\
         &$N_{so}$& $n_{occ}^{\alpha}$& $n_{occ}^{\beta}$& $n_{vir}^{\alpha}$& $n_{vir}^{\beta}$ & \tbc{1}{(GB)}       &  \tbc{1}{(GB)}  & \\
\hline
Uracil           &760& 29& 28&351  &352   &3$\times$4.5&26.2& 2044 \\
ESCA             &880& 36& 35& 404& 405    &3$\times$ 9.2&40.6 &2423 \\
Benzene-Ammonia  &940& 26& 25& 444& 445   &3$\times$5.8 &49.5&2457 \\ 
Zn-porphyrin     &2450& 95& 94&1130&1131&3$\times$ 510.2&876.6&6258\\ 
\hline  
\end{tabular}\\
\end{table*}


\subsection{Geometries, Basis Sets, and Computational Details}

In this study, we perform RT-EOM-CC simulations of two molecules, formaldehyde and ethyl trifluoroacetate (ESCA), in order to compare to previous theoretical and experimental results. For formaldehyde, we used the experimental geometry\cite{NIST} and studied both C and O core ionizations using either the aug-cc-pVDZ,\cite{aug-cc-pvdz} aug-cc-pVTZ,\cite{aug-cc-pvdz} or Sapporo-TZP\cite{sapporo-tzp_1, sapporo-tzp_2} basis sets for all atoms in the molecule.
Given that the ESCA molecule contains four C atoms, we computed ionization spectra from each  of them. These calculations were performed with the Sapporo-TZP basis set for all the first-row elements, while for the H atom we used the aug-cc-pVDZ basis set. Since a complete experimental geometry of the ESCA molecule is not available, we used the one obtained from a B3LYP/aug-cc-pVTZ optimization.\cite{esca_geometry}
The real-time time-propagation used a time-step of 0.015 au ($\sim$0.36 as) for formaldehyde and 0.01 au ($\sim$0.24 as) for the ESCA molecule, with a convergence cutoff of $10^{-4}$ for the fixed-point micro-iteration solution of the implicit first-order Adams--Moulton integrator. The total propagation time was 450 au ($\sim$11 fs) for formaldehyde and 100 au ($\sim$2.5 fs) for ESCA.
All stationary calculations were performed with a linear dependence threshold for the basis sets 
of 10$^{-6}$, SCF convergence cutoff of 10$^{-8}$ au for the energy, Cholesky diagonal cutoff of 10$^{-6}$ and a CCSD convergence cutoff of 10$^{-8}$ au. None of the virtual orbitals are frozen in any of our simulations.

\section{Results and Discussion}
\subsection{Performance Analysis}

\begin{figure}[t]
\includegraphics[clip,width=0.49\textwidth]{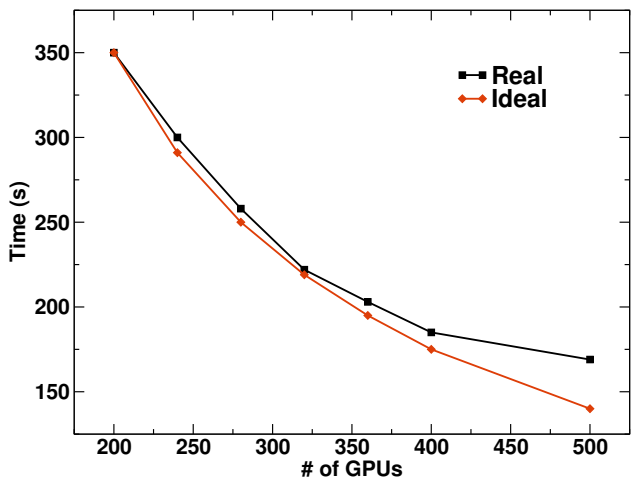}
\caption{Real \textit{vs} ideal single micro-iteration time as a function of number of GPUs for Zn-porphyrin using 655 basis functions.}
\label{fig:strong_scaling}
\end{figure}

As described in  previous sections, a simulation is divided into a series of ``macro-iterations'' associated with each time step, and within each macro-iteration, many ``micro-iterations'' are performed to solve Eq. \ref{eq:trapzrule}. Thus, the fundamental parallel performance metric for RT-EOM-CC is the ``\textit{time per micro-iteration}'' associated with the calculation of the matrix elements in Eq. \ref{eq:trapzrule}.
In order to investigate the performance and scalability of the new optimized TAMM implementation of RT-EOM-CCSD, we have performed a series of calculations on the ``pre-production'' NERSC Perlmutter system (Nvidia A100).
If  the scaling is ideal, the  total compute time is inversely proportional to the total number of allocated processors provided the total number of mathematical operations in each test case remains the same.
Figure \ref{fig:strong_scaling} shows the scalability of the TAMM implementation of RT-EOM-CCSD for Zn-porphyrin using the aug-cc-pVDZ basis set, which results in a total of 655 basis functions ($n_{occ}$=94, $n_{vir}$=561) after pruning 122 linear dependencies. Our time-dependent simulations involve a configuration space of 
$n_{occ}^{\alpha}$ = 95, $n_{vir}^{\alpha}$ = 560, $n_{occ}^{\beta}$ = 94, $n_{vir}^{\beta}$ = 561. We explore scaling between 200 and 500 GPUs (50 to 125 nodes given that each node is connected to 4 GPUs), using the performance with 200 GPUs as the reference. Figure \ref{fig:strong_scaling} also shows the ideal theoretical scaling.
In this range, we observe a nearly ideal drop in the computation time for each micro-iteration, very close to the theoretically limit.
The parallel efficiency remains high (i.e. higher than 94\%) up to 400 GPUs, after which it drops to 83\% for 500 GPUs.





\begin{figure}[t]
\includegraphics[clip,width=0.49\textwidth]{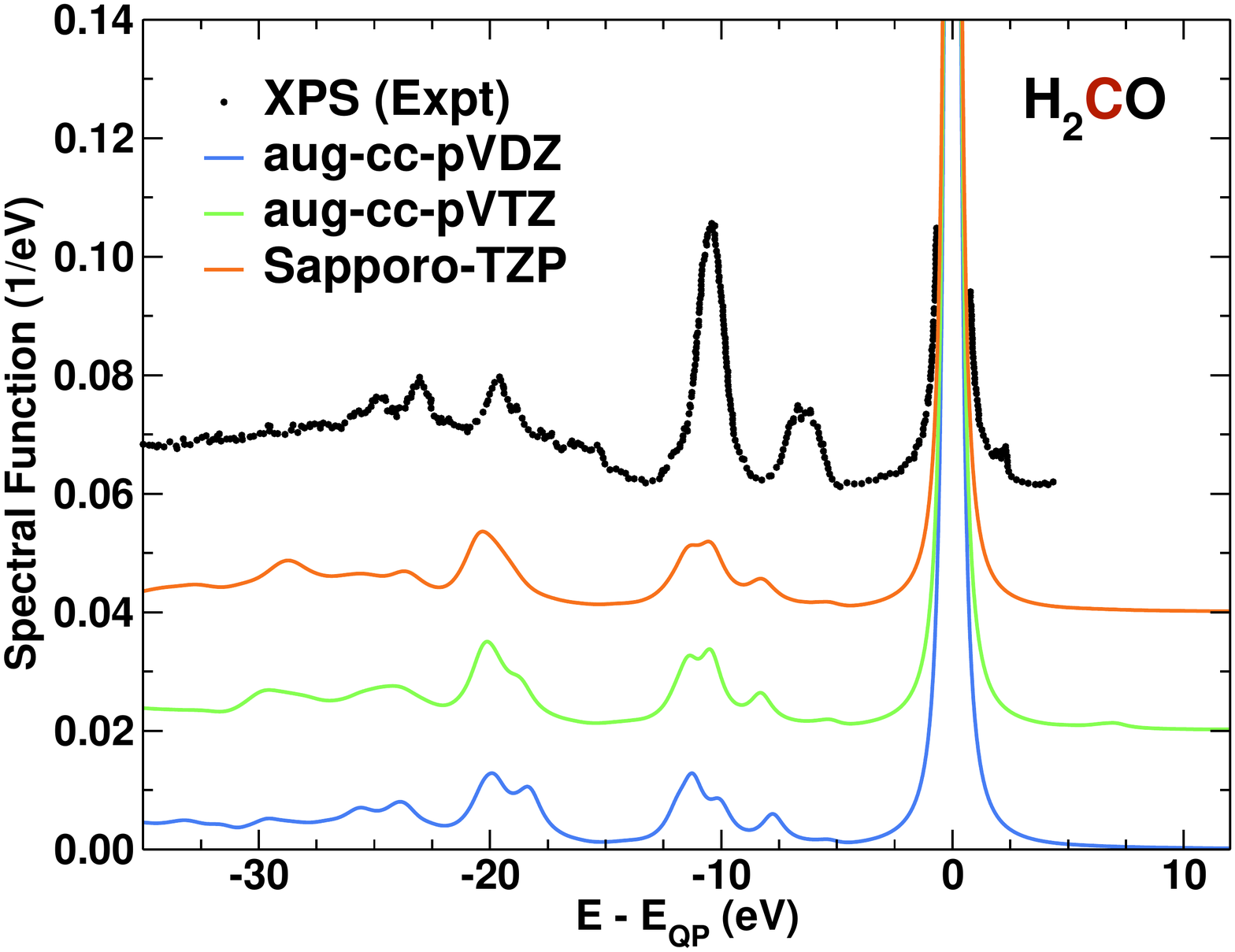}
\includegraphics[clip,width=0.49\textwidth]{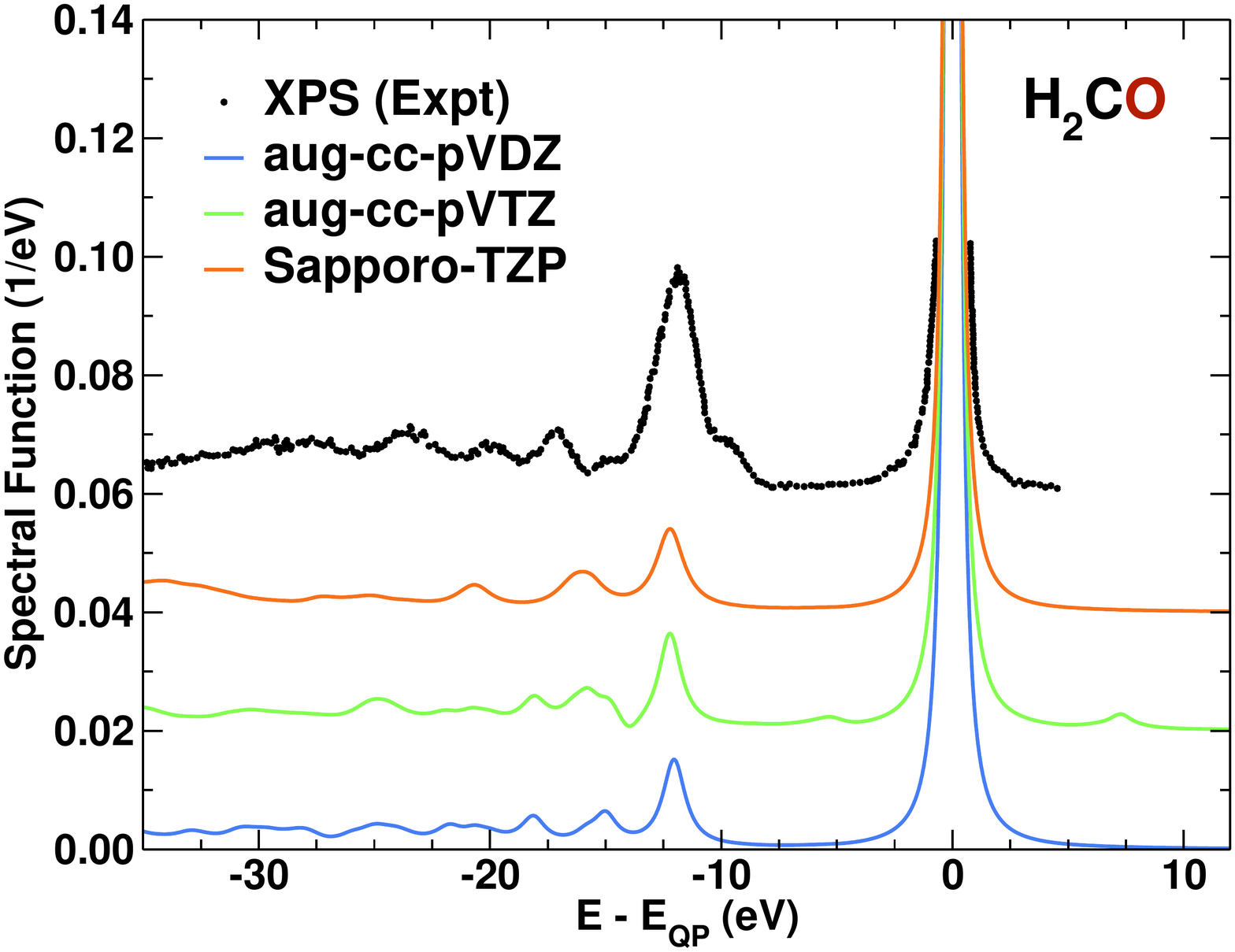}
\caption{Comparison of satellite regions of the C (top) and O (bottom) RT-EOM-CCSD core spectral functions to the XPS experiment\cite{KURAMOTO2005253} for formaldehyde (H$_2$CO) as a function of basis set. The data has been shifted so that the quasiparticle peak is at 0 eV.}
\label{fig:SPF_H2CO}
\end{figure}

\subsection{Spectral Function Results}
The high quality of the RT-EOM-CC results has been previously demonstrated\cite{vila2021equation,vila2022_water}. Thus here we focus  on showcasing the new capabilities of the RT-EOM-CCSD TAMM implementation for more complex systems. For this purpose, we have simulated formaldehyde (H$_2$CO) and ethyl trifluoroacetate (ESCA). For formaldehyde, we focus on the satellite region of the C and O core spectral functions (Fig. \ref{fig:SPF_H2CO}). We find that the theory reproduces the XPS semi-quantitatively, with the position of most of the satellite features relative to the quasiparticle peak in reasonable agreement with experiment. The relative intensities of the peaks are not as well reproduced, probably due to i) the limitations of the local valence basis set used that misses the continuum background contribution, and ii) the limitations of the RT-EOM-CCSD to include all the relevant excitations in this energy range.

\begin{figure}[t]
\includegraphics[clip,width=0.49\textwidth]{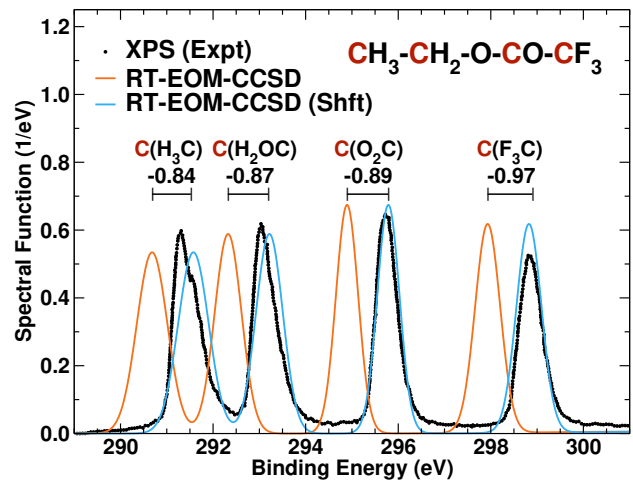}
\caption{Comparison of the RT-EOM-CCSD/Sapporo-TZP C core spectral functions (red) to the experimental\cite{esca_geometry} XPS (black dots) for the ethyl trifluoroacetate (ESCA) molecule. Each peak corresponds to an individual C core ionization and has been broadened to match the vibrational experimental broadening. Also shown are the same results shifted by 0.89 eV (blue).}
\label{fig:SPF_ESCA}
\end{figure}

For the case of the ESCA molecule, shown in Fig. \ref{fig:SPF_ESCA}, we calculated the C core spectral function for each of the inequivalent C atoms in the system. The agreement with experiment is excellent, apart from a nearly-constant overall underestimation of the binding energies. When the underestimation shift is removed, the relative mean absolute error is only 0.04 eV. We also find that, unlike for the satellite peaks in H$_2$CO, the relative intensities of the ESCA quasiparticle peaks are qualitatively reproduced by the theory. By fitting the experiment to a skew Gaussian distribution (to account for the vibrational asymmetry), we find that the intensities of the C(H$_2$OC), C(O$_2$C) and C(F$_3$C) peaks relative to the C(H$_3$C) one are 0.81, 0.90 and 0.91, respectively, while for the theory the ratios are 0.94, 0.91 and 0.96.
It was speculated\cite{esca_geometry} that some of the intensity of the different C quasiparticle peaks might originate from underlying satellite peaks from lower energy cores. We find that for each of the individual spectral functions the first satellite peaks appear more than 10 eV above the quasiparticle, and thus all the intensity observed for the peaks between 291 and 299 eV can be assigned exclusively to the quasiparticle transitions.

\section{Conclusions}

We have successfully implemented the RT-EOM-CCSD method within the parallel TAMM infrastructure using Cholesky decomposed two-body repulsion matrix elements. This implementation eliminates the memory bottleneck of the original approach associated with storing two-electron integrals.
Unlike our earlier TCE-based RT-EOM-CCSD implementation, which relied only on real algebra, our new TAMM implementation supports explicit complex algebra. This implementation is also flexible regarding the choice of the reference function (i.e. where the hole state is located), and has checkpointing/restart capabilities at any stage of the workflow. This is quite important since the propagation portion of the workflow can be very time-consuming and restarts are usually needed in shared computing systems.
Moreover, the TAMM RT-EOM-CCSD shows very good scalability, paving the way for simulations of larger, more realistic and chemically relevant systems, employing larger basis sets in conjunction with reduced memory requirements. Illustrative calculations for the formaldehyde and ESCA molecules  demonstrate that the predicted positions for the quasiparticle and satellite peaks are in good agreement with experimental values. In particular, the RT-EOM-CCSD reproduced the relative position of the different core ionizations in the ESCA molecule, highlighting its capabilities to study chemical speciation. The method describes the positions of the satellite peaks without any corrections to the quasiparticle-satellite gap.


Finally, work is in progress on an atomic orbital-based implementation, coupled to more efficient solvers for the implicit Adams--Moulton propagator that should reduce the computational time by at least an order of magnitude. Other future methodological developments include the implementation of spin-orbit coupling, and the multi-component coupled-cluster formalism. These extensions will allow first-principles studies of multielectron dynamics in previously unreachable large chemical systems, such as simulations involving multiple core holes.

\begin{acknowledgement}
This work was supported by the Computational Chemical Sciences Program of the U.S. Department of Energy, Office of Science, BES, Chemical Sciences, Geosciences and Biosciences Division in the Center for Scalable and Predictive methods for Excitations and Correlated phenomena (SPEC) at PNNL, with computational support from NERSC, a DOE Office of Science User Facility, under contract no. DE-AC02-05CH11231. B.P. also acknowledges support from the Laboratory Directed Research and Development (LDRD) Program at PNNL.
\end{acknowledgement}


\bibliography{achemso-demo}
\end{document}